\pdfoutput=1
\documentclass[twocolumn,aps,pra,floatfix,superscriptaddress]{revtex4-2}
\usepackage{amsmath,amssymb,physics,bm}
\usepackage{graphicx,float}
\usepackage{times,txfonts}
\usepackage{color}
\usepackage{multirow}
\usepackage{bbold}
\usepackage{color}
\usepackage{soul,xcolor}

\usepackage[colorlinks=true,linkcolor=blue,citecolor=blue,urlcolor =blue]{hyperref}
\usepackage{times,txfonts}
\usepackage{float}
\usepackage{natbib}
\usepackage{blindtext}
\usepackage[export]{adjustbox}
\usepackage{mathrsfs}
\usepackage{textcomp}
\usepackage{blkarray}
\usepackage{multirow}

\newcommand{\iu}{\mathrm{i}}
\newcommand{\eu}{\mathrm{e}}
\setstcolor{red}

\begin{document}
\title{Quantum control of Rydberg atoms for mesoscopic quantum state and circuit preparation }

\author{Valerio Crescimanna}
\thanks{These authors contributed equally to this work}
\affiliation{National Research Council of Canada, 100 Sussex Drive, Ottawa, Ontario K1N 5A2, Canada}
\affiliation{Department of Physics, University of Ottawa, 25 Templeton Street, Ottawa, Ontario, K1N 6N5 Canada}
\author{Jacob Taylor}
\thanks{These authors contributed equally to this work}
\affiliation{National Research Council of Canada, 100 Sussex Drive, Ottawa, Ontario K1N 5A2, Canada}
\affiliation{Institute for Quantum Computing,
University of Waterloo,
200 University Avenue West,
Waterloo, Ontario  N2L 3G1, Canada}
\author{Aaron Z. Goldberg}
\affiliation{National Research Council of Canada, 100 Sussex Drive, Ottawa, Ontario K1N 5A2, Canada}
\affiliation{Department of Physics, University of Ottawa, 25 Templeton Street, Ottawa, Ontario, K1N 6N5 Canada}
\author{Khabat Heshami}
\affiliation{National Research Council of Canada, 100 Sussex Drive, Ottawa, Ontario K1N 5A2, Canada}
\affiliation{Department of Physics, University of Ottawa, 25 Templeton Street, Ottawa, Ontario, K1N 6N5 Canada}
\affiliation{Institute for Quantum Science and Technology, Department of Physics and Astronomy, University of Calgary, Alberta T2N 1N4, Canada}

\begin{abstract}
Individually trapped Rydberg atoms show significant promise as a platform for scalable quantum simulation and for development of programmable quantum computers. In particular, the Rydberg blockade effect can be used to facilitate both fast qubit-qubit interactions and long coherence times via low-lying electronic states encoding the physical qubits. To bring existing Rydberg-atom-based platforms a step closer to fault-tolerant quantum computation, we demonstrate high-fidelity state and circuit preparation in a system of five atoms. We specifically show that quantum control can be used to reliably generate fully connected cluster states and to simulate the error-correction encoding circuit based on the ‘Perfect Quantum Error Correcting Code’ by Laflamme \textit{et al}. [Phys. Rev. Lett. 77, 198 (1996)]. Our results make these ideas and their implementation directly accessible to experiments and demonstrate a promising level of noise tolerance with respect to experimental errors. With this approach, we motivate the application of quantum control in small subsystems in combination with the standard gate-based quantum circuits for direct and high-fidelity implementation of few-qubit modules.
\end{abstract}

\maketitle

\section{Introduction}
Quantum computing represents a paradigm shift in information processing. A scalable and fault-tolerant quantum computer will likely offer computational advantage in particular areas. Arguably, gate-based quantum computation and, to some extent, measurement-based quantum computation~\cite{nielsen2002quantum,raussendorf2001one,briegel2009measurement} are the current dominant paths forward for the leading implementations of quantum computing~\cite{acharya2022suppressing,krinner2022realizing,benhelm2008towards,postler2022demonstration,PhysRevX.12.021049,PhysRevX.11.041058,bourassa2021blueprint}. Unlocking the full potential of quantum computation will depend on our ability to resolve barriers to scalability and fault-tolerance. Quantum control is effective in optimizing the performance of individual gates~\cite{RevModPhys.76.1037,zahedinejad2015high,ball2021software}. In this work, we extend the application of quantum control to high-fidelity generation of multi-qubit quantum states and circuits where sequential application of multiple multi-qubit gates is necessary. 

Our approach to extend quantum control to multiple qubits is based on unitary constructions inspired by the quantum approximate optimization algorithm (QAOA)~\cite{farhi2014quantum}. The QAOA, by Farhi {\it et al.}, proposes an approach to find approximate solutions to combinatorial optimization problems that are achieved by a Trotterized approximation of the quantum adiabatic dynamics \cite{farhi2014quantum}. This has been shown to be universal for quantum computation with Hamiltonians that are homogeneous sums of single-qubit Pauli X and two-local ZZ operators \cite{lloyd2018quantum}. One necessity of quantum technologies is to prepare specific quantum states or enact specific unitary operations. This should always be possible, but may require prohibitively complex multi-qubit quantum gates. Here, instead, we apply quantum control to realize desired states and circuits through an algorithm that requires interactions among qubits that are at most two-local and are restricted to a linear geometry. 

Rydberg atoms are a particularly attractive system for implementing these ideas. With the remarkable progress in development of arrays of individually trapped Rydberg atoms and their application in quantum simulation \cite{Bernien2017,keesling2019quantum,Browaeys2020,ebadi2021quantum,scholl2021quantum,taylor2022simulation}, it is desirable to enhance our abilities in quantum computation with Rydberg atoms. In fact, Rydberg atoms can be aligned today with optical tweezers to form an array of tens of atoms, making this one of the largest platforms for physically realizing quantum computers in terms of number of qubits \cite{ebadi2021quantum}. The local gates controlling the single atoms can be realized through Rabi oscillations by properly setting the frequency, intensity, and exposure times to laser pulses \cite{PhysRevA.85.042310}. The strong long-range interaction that leads to the so-called Rydberg blockade \cite{saffman2010quantum,vsibalic2018rydberg} is activated by instead exciting atoms to their high-lying Rydberg states. In such a way, all the operators required by our algorithm can be experimentally realized with the optimization variables that are mapped accordingly to tunable physical parameters of the lasers~\cite{PhysRevLett.85.2208, RevModPhys.82.2313, PhysRevA.93.012306}. 

Thus far, Rydberg atoms have been used for generating and detecting non-trivial quantum states of many atoms~\cite{omran2019generation,bluvstein2022quantum,madjarov2020high} based on engineering the energy spectrum of a Rydberg-atom quantum simulator. As some quantum states remain far from energy eigenstates of any local Hamiltonian~\cite{PhysRevLett.91.210401}, it is important to consider methods such as ours for generation of arbitrary quantum states or simulation of quantum circuits beyond a single operation.

We begin with an introduction to QAOA, a comparison with quantum annealing (QA), and a description of the universal quantum control developed from the optimization algorithm. We discuss the mapping of the parameters used for the implementation with the Rydberg atoms in the Section \ref{sec:equivalence}. In Section \ref{sec: state and circuit}, we introduce state preparation and circuit simulation examples to test our protocol. We then evaluate the performance of the algorithm for different implementation depths and noise levels in Section \ref{sec:results}, providing an important perspective on resource trade-offs for quantum control in realistic conditions.

\section{The QAOA}
The QAOA is a hybrid classical-quantum variational algorithm able to provide approximate solutions to optimization problems. QAOA was originally introduced to address the MaxCut problem \cite{farhi2014quantum}, an important combinatorial problem in graph theory. In turn, the MaxCut problem can be encoded into the Sherrington-Kirkpatrick model for spin glasses by tuning the interactions among a series of spins. Thanks to this equivalence, finding the ground state of a spin-glass Hamiltonian is equivalent to finding the solution of the MaxCut problem encoded within. In a similar way, whenever an analogous encoding into a physical system can be found for a combinatorial problem under investigation, the solution corresponds to the ground state of the Hamiltonian of the system. QAOA then guarantees to find the ground state of such a Hamiltonian with a fidelity that increases with the number of steps chosen for the algorithm. 

The QAOA can be interpretted as an evolved version of the well-known QA algorithm. With QA, the initial state is set to be the ground state of an initial Hamiltonian, which is the so-called mixing Hamiltonian. Then the state is let to evolve under a time-dependent Hamiltonian that converges, after a sufficiently long time, to the cost Hamiltonian whose ground state is sought. This scheme was originally proposed to solve the Ising problem \cite{PhysRevE.58.5355}, where the Ising model is a well-known specialization of the Sherrington-Kirkpatrick model, which was conveniently used to encode the MaxCut problem and has a broad range of additional applications \cite{Verresen2023arxiv}.

A system of $N$ spins can have couplings with any number of the other spins arbitrarily far away and will have its energy affected by an external magnetic field $\bf{h}$. If the direction of this field is chosen to be the $z$-axes of the spins, the resulting generalized Sherrington-Kirkpatrick model has Hamiltonian \cite{Panchenko2012}
 \begin{equation}\label{eq:SKHamiltonian}
         H_Z=\sum_{i_1}^Nh_{i_1}Z_{i_1}+\sum_{i_1,i_2}^NJ_{i_1i_2}Z_{i_1}Z_{i_2}
          +\sum_{i_1,i_2,i_3}^NJ_{i_1i_2i_3}Z_{i_1}Z_{i_2}Z_{i_3}+\cdots \, .
 \end{equation}
Here, $J_{i\cdots k}$ is the $k$-local interaction energy between the qubits at sites $i,\cdots, k$ and $Z_{i}$ represents the Pauli operator $\sigma_z$ acting on the $i$th qubit of the $N$-qubit state. Different parameters $J_{i\cdots k}$ correspond to different physical situations and also to different combinatorial problems.

In order to identify the ground state of the cost Hamiltonian \eqref{eq:SKHamiltonian}, QA proposes to initialize an $N$-qubit state that is a linear superposition of all of the computational basis states. This is chosen to be an easy-to-prepare eigenstate of the mixing Hamiltonian, where the latter is named because it combines all of the basis states. An example for such a state is $\ket{++\cdots+}$, which is an eigenstate of the Hamiltonian $H_X=\sum_i X_i$ with eigenvalue $+N$. Thanks to this initialization, any eigenstate of the cost Hamiltonian $H_Z$ can be reached when the state evolves under a time-dependent Hamiltonian that starts from $H_X$ and equals the cost Hamiltonian after a sufficiently long time $\tau$. 
The Hamiltonian used to evolve the state in the QA is a time-dependent, linear combination of the cost Hamiltonian $H_Z$ and the mixing Hamiltonian $H_X$:
\begin{equation}\label{eq:QAHamiltonian}
    H(t)=\left(1-\frac{t}{\tau}\right)H_X+\frac{t}{\tau}H_Z.
\end{equation}
Since the cost and the mixing Hamiltonian do not commute with each other, realizing the evolution under each of the two terms independently requires using the Suzuki-Trotter decomposition into a formally infinite number of alternating steps of applying $H_X$ and $H_Z$: $\exp(A+B)=\lim_{p\to\infty}\left(\eu^{A/p}\eu^{B/p}\right)^p$.
 
The probability of evolving to the actual ground state of the cost Hamiltonian $H_Z$ improves for larger values of the coefficient $\tau$ that defines the temporal dependence of the Hamiltonian. However, letting the system evolve for too long times tends to expose it to decoherence. QAOA partially intervenes to solve this kind of problem. 
In QAOA, the Suzuki-Trotter formula is used to separate the non-commuting terms that form the time-dependent Hamiltonian of QA, the mixing Hamiltonian and the cost Hamiltonian, but with the caveat that each is only applied a finite number of times. The QAOA thus ends up being an algorithm with $p$ steps where the unitary evolutions under the two Hamiltonians can be more easily realized, since they are time-independent and can be encoded in quantum gates. 

In fact, the Trotterization of QA with a finite number of steps $p$ directly approximates the evolution that one would have with the time-dependent Hamiltonian required by QA. However, each of the time parameters of the QAOA can be classically optimized to provide a state whose fidelity is larger than what QA might provide after annealing for the same amount of time. The optimal state is then found by alternatively applying the unitary evolutions that describe the evolution under $H_X$ and $H_Z$: 
\begin{equation}\label{eq:evolutionQAOA}
    \ket{\psi_f}=\exp(-\iu\alpha_pH_X)\exp(-\iu\beta_p H_Z)\cdots\exp(-\iu\beta_1 H_Z)\ket{\psi_i},
\end{equation}
where the parameters $\alpha_i$, and $\beta_i$ are the so-called Rabi angles at the $i$th step.
To wit, the QAOA can be read as a Trotterization of an adiabatic schedule with a finite number of iterations in which the time parameters are tuned through a classical optimization.
 
Moreover, the QAOA is a universal algorithm \cite{lloyd2018quantum}. In fact, if the mixing Hamiltonian $H_X$ and the cost Hamiltonian $H_Z$ are set to be
\begin{equation}\label{eq:HXHZ}
    \begin{aligned}
    H_X&=\sum_iX_i ,\\
    H_Z&=\sum_{i}\beta^{(0)} Z_{2i}+\beta^{(1)} Z_{2i+1}+\gamma^{(0)} Z_{2i}Z_{2i+1}+\gamma^{(1)} Z_{2i+1} Z_{2i}\\
    &\equiv \beta^{(0)} H_Z^{(0)}+\beta^{(1)} H_Z^{(1)}+\gamma^{(0)} H_{ZZ}^{(0)}+\gamma^{(1)} H_{ZZ}^{(1)},
    \end{aligned}
\end{equation} the algorithm is capable of reaching any desired state. In the above equations, the sum is made over all of the qubits of the $N$-qubit state and $X_i$ is the $\sigma_x$ Pauli operator applied to the $i$th qubit. In this case, Eq.~\eqref{eq:evolutionQAOA} reads 
\begin{equation}\label{eq:evolutionQAOA2}
\begin{split}
   \ket{\psi_f}=\eu^{-\iu\alpha_nH_X}\eu^{-\iu\beta_n^{(0)}H_Z^{(0)}}\eu^{-\iu\beta_n^{(1)}H_Z^{(1)}}\eu^{-\iu\gamma_n^{(0)}H_{ZZ}^{(0)}}\eu^{-\iu\gamma_n^{(1)}H_{ZZ}^{(1)}}\\
\quad \cdots\times \eu^{-\iu\alpha_1H_X}\eu^{-\iu\beta_1^{(0)}H_Z^{(0)}}\eu^{-\iu\beta_1^{(1)}H_Z^{(1)}}\eu^{-\iu\gamma_1^{(0)}H_{ZZ}^{(0)}}\eu^{-\iu\gamma_1^{(1)}H_{ZZ}^{(1)}}\ket{\psi_i} .
\end{split}
\end{equation}

All of the terms in the Hamiltonian have the advantage of being at most two-local, so they do not require physical systems with long range interactions. 
Since all the terms in the cost Hamiltonian commute, the Baker-Campbell-Hausdorff formula has been used to write the exponential of the sum as a product of exponentials, given that the individual exponentials can be more easily encoded in a gate-based circuit. Despite the just described similarity between QAOA and a trotterization of QA, there exists a deep difference between the two protocols. With QA, the ground state of a given cost Hamiltonian is obtained by evolving the initial state under the time-dependent Hamiltonian $H(t)$ of Eq.~\eqref{eq:QAHamiltonian}. In fact, given a cost Hamiltonian $H_Z$, and an initial state $\ket{\psi_i}$ the resulting state $ \ket{\psi} = \int_0^\tau e^{-iH(t)}\ket{\psi_i}\text{d}t$, should minimize $\expval{H_Z}{\psi}$ for sufficiently large values of $\tau$. The optimization of the parameters $\alpha,$ and $\bm{\beta}$ in Eq.~\eqref{eq:evolutionQAOA2} can be performed in a similar fashion to generate an evolution that minimize  $\expval{H_Z}{\psi_f}$, where $\ket{\psi_f}$ is as defined in \eqref{eq:evolutionQAOA2}. However, the protocol is much more general than this, and the parameters $(\alpha,\bm{\beta})$ can be used to  find the ground state of any cost Hamiltonian $H_C$ by minimizing $\expval{H_C}{\psi_f}$, or even quantum states $\ket{\psi_T}$ that are far from ground state of any local Hamiltonian if the cost function to minimize is set to be $-\abs{\braket{\psi_T}{\psi_f}}^2$. Analogously, with QAOA, we can approximate with arbitrary precision an target operator $O$. With regard to these last two applications, the terms $H_X$ and $H_Z$ of Eq.~\eqref{eq:HXHZ} lose the role of mixing and cost Hamiltonian respectively that has been given to them in the context of QA, but the evolution of the states under these Hamiltonians remains functional for QAOA. In summary, QAOA can be used to realize any state with arbitrary precision by only employing Hamiltonians that are at most two-local. Using longer sequences allows the creation of more complicated states whose entanglement can grow in a known fashion \cite{Dupontetal2022}. The algorithm then provides a recipe for the realization of any quantum state in the chosen physical implementation. 
This has an advantage over other generation protocols that rely on extra resources such as ancillary qubits, perfect local operations, and Bell-state measurements~\cite{PhysRevLett.127.220503}; our algorithm has the benefit of being able to incorporate realistic experimental imperfections.

By using the QAOA protocol, we investigate the generation of a variety of states including Absolutely Maximally Entangled states with 5 and 6 qubits and GHZ states, all of which can successfully be generated by the protocol. Some of those specific states can be created with specific protocols, but ours can be applied to the preparation of arbitrary states and circuits. We focus our in-depth discussion on fully connected Cluster states, and  Error-Correction Encoding Circuit, which are challenging to produce with alternative approaches.  
As an example, in a Rydberg platform, an error encoding circuit might be more conveniently generated if long range interactions beyond nearest neighbours can be introduced; however, this would come a the cost of larger errors proportional to the number of sites involved in the interaction itself~\cite{Isenhower2011}, which is why we seek alternative strategies relying only on simpler interactions.

\section{Rydberg Scheme}
Rydberg states are particular excited states of an electron of an atom that is subjected to a Coulomb potential \cite{SCHAFER200935}.  
Those states are usually realized with atoms of alkali elements such as potassium, cesium, and, most often, rubidium. In the alkali elements, the electron farthest from the nucleus is shielded from the latter's attractive potential by all the other electrons that completely fill the lower energetic levels of the same atom. The less intense effective potential leads to easier control of the outermost electron and, as a result, Rydberg atoms behave in many respects like hydrogen atoms and can be described accordingly for several applications. 

\begin{figure}[ht]
\includegraphics[viewport=100 480 600 700 ,clip,width=0.5\textwidth]{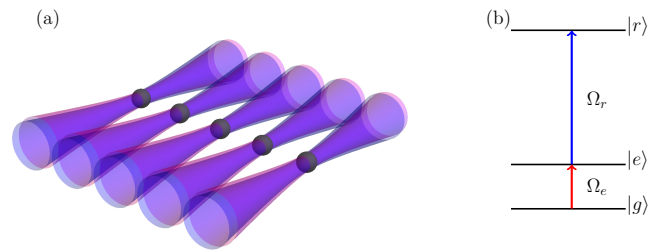}    
        \caption{(a) Dipole trapped Rydberg atoms within a line separated by a distance of $5.74\, \mu \mathrm{m}$ used for the purpose of generating non-trivial states, with inlay of Rydberg level structure. (b) The red arrow represents the global driving pulse, driving $\ket{g} \mapsto \ket{e}$, the blue arrow represents the atom-specific resonant driving pulse from $\ket{e} \mapsto \ket{r}$, and dipole trapping consists of a third set of lasers perpendicular to the plane.}\label{fig:RydbergAtoms}
\end{figure}

Rydberg atoms can be collected in arrays using dipole traps like the optical tweezers depicted in Fig. \ref{fig:RydbergAtoms}(a). For each atom, we define three accessible energetic levels (Fig.~\ref{fig:RydbergAtoms}(b)). The ground state $\ket{g}\equiv\ket{0}$ and the excited state $\ket{e}\equiv\ket{1}$ are set to be the states defining the computational basis. Furthermore, the electron may accede to the Rydberg state, a third energetic level $\ket{r}$ with larger energy. 

The Rydberg state allows the realization of an entangling gate between a pair of Rydberg atoms thanks to the long-range interaction that arises when one atom is on this last energetic level; this phenomenon is known as the Rydberg blockade. 
In the Rydberg blockade, one atom in the Rydberg state acts as the control by shifting the same energetic level of the target atom through Van der Waals forces, due to a specific dipole interaction between the atoms that increases with the distance of the electron from the nucleus.
The Hamiltonian that describes the interactions of the Rydberg atoms with themselves and a shining laser is 
\begin{equation}
    \frac{H}{\hbar}= \sum_{i}  \frac{\Omega_i(t)}{2} (\sigma_i^{1 \mapsto r}+\sigma_i^{r \mapsto 1})-\sum_{i} \Delta(t) n_i +\sum_{i>j} V_{ij} n_i n_j ,
\end{equation}
where $i$ ranges over all of the atoms, $\sigma_i^{m \mapsto n}=|n\rangle_i\langle m|$ sends the $i$th atom in state $\ket{m}$ to state $\ket{n}$, and $n_i=\ket{r}_i\bra{r}$, while $\Omega$ is the Rabi frequency, $\Delta$ is the detuning, and $V$ labels the so-called van der Waals interaction that scales with the sixth power of the distance between the atoms~\cite{Bernien2017}. The influence of this interaction extends up to the so-called blockade radius, so it can be switched off for non-nearest-neighbor qubits with a suitable choice of the spacing between atoms~\cite{vsibalic2018rydberg}.
 
When the control atom is in the Rydberg state and the target atom in the excited state, a $2\pi$-pulse on the target does not lead to any change of phase; whereas, such a pulse would flip the phase if the control atom was in the ground state. This behaviour allows us to implement the two-qubit logical gate required by QAOA. 
\section{Rydberg equivalence}
\label{sec:equivalence}
In order to experimentally realize the single-qubit operators on the dipole-trapped Rydberg atoms, the Rabi cycle is used. In the rotating-wave approximation and when the frequency of the laser is resonant with the energy splitting between the two logical states, the Hamiltonian for a particular atom is given by \begin{equation}
    H_x=\Omega\sigma_x ,
\end{equation} where $\Omega$ is one half of the Rabi frequency of the system, and $\sigma_x$ is the NOT or bit-flip quantum gate that acts on one qubit. As a consequence, the mapping between the time parameter $\alpha$ introduced in Eq.~\eqref{eq:evolutionQAOA2} and the pulse time of the laser is fixed by $\alpha=\Omega t$; the same pulse must act on all $N$ qubits to enact $H_X$. 
  
The second operator that is used to realize universal quantum control with QAOA is the single qubit operator $\sigma_z$. The unitary operator that describes the evolution under the Hamiltonian for a two-level atom driven by a coherent laser with detuning $\Delta$ and generalized Rabi frequency $\Omega^\prime =\sqrt{\Omega^2+\Delta^2}$ is
\begin{widetext}
 \begin{equation}\label{eq:timeEvolution}
    U(t)=\begin{pmatrix} 
    \eu^{\frac{\iu\Delta t}{2}}\left(\cos\left(\frac{\Omega' t}{2}\right)-\iu\frac{\Delta}{\Omega'}\sin\left(\frac{\Omega' t}{2}\right)\right) && \iu\eu^{\frac{\iu\Delta t}{2}}\frac{\Omega}{\Omega'}\sin\left(\frac{\Omega' t}{2}\right)\\
    \iu\eu^{-\frac{\iu\Delta t}{2}}\frac{\Omega}{\Omega'}\sin\left(\frac{\Omega' t}{2}\right) && \eu^{-\frac{\iu\Delta t}{2}}\left(\cos\left(\frac{\Omega' t}{2}\right)+\iu\frac{\Delta}{\Omega'}\sin\left(\frac{\Omega' t}{2}\right)\right).
    \end{pmatrix}
\end{equation}
\end{widetext}
 When the time is $t=\frac{2\pi}{\Omega^\prime}$ and the detuning $\Delta$ is fixed to $\Delta=-\frac{\beta\Omega}{\sqrt{\pi^2-\beta^2}}$, the evolution of Eq. \eqref{eq:timeEvolution} corresponds to evolution under the Hamiltonian $Z$ for a particular qubit: $U_Z(t)=\exp(-\iu\beta Z)$. Acting with the same gate on the even (odd) qubits with a detuning fixed by $\beta^{(0)}$ ($\beta^{(1)}$) enacts $H_Z^{(0)}$ ($H_Z^{(1)}$).
 
\begin{figure}
\includegraphics[trim=0 0 0 0 ,clip,width=0.4\textwidth]{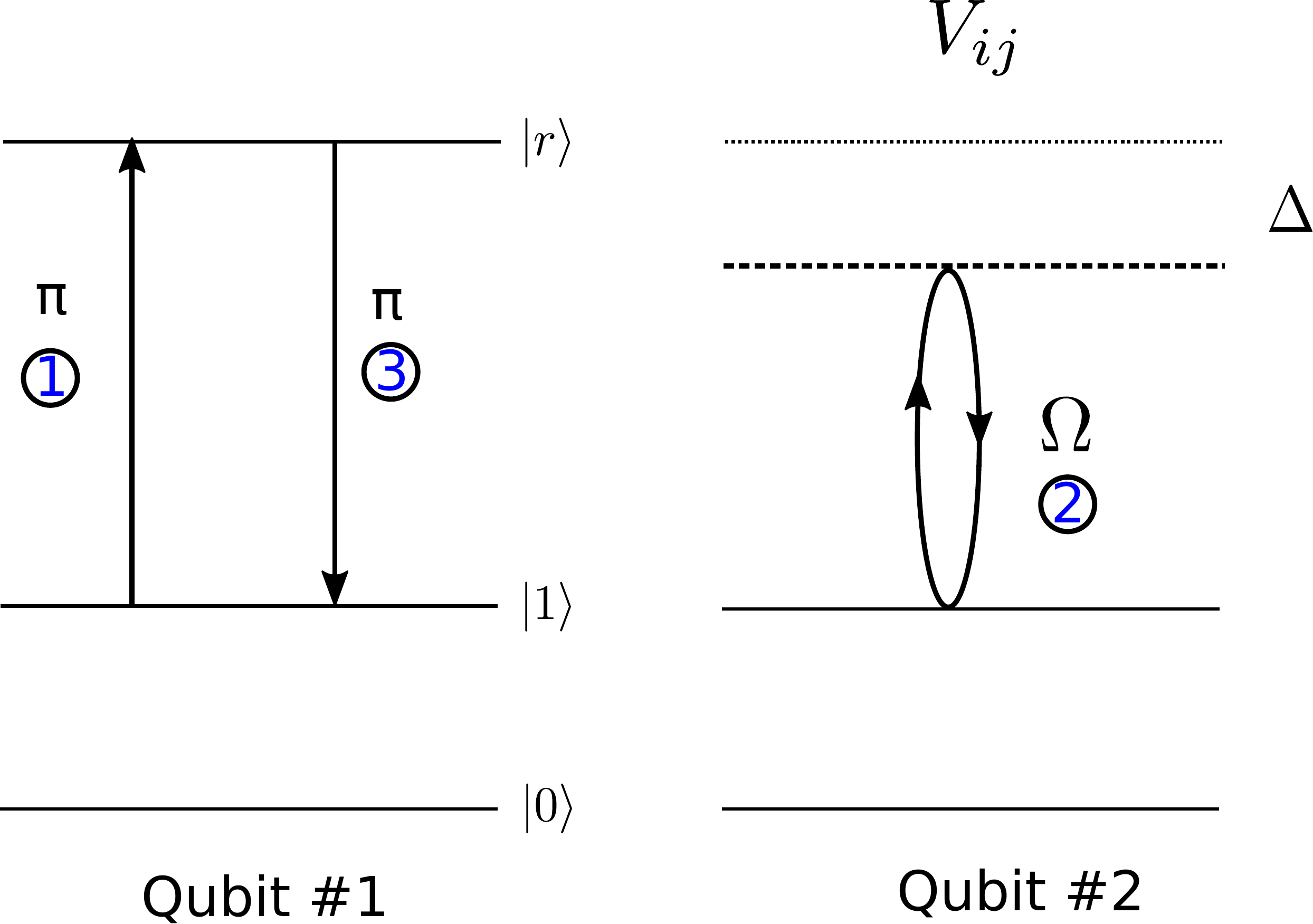}
\caption{ Formulation of $H_{ZZ}$ unitary operator in the Rydberg system: Consists of 2 Rydberg atoms and 3 pulses. We have updated the labels $\ket{g}$ and $\ket{e}$ to be $\ket{0}$ and $\ket{1}$ to enforce their role as logical qubit states.}
\label{fig:HzzDiagram}
\end{figure}

Finally, we want to implement a two-qubit entangling operator. The scheme to realize a controlled-$Z$ gate consists of three steps as depicted in Fig.~\ref{fig:HzzDiagram}. 
First, a $\pi$-pulse is applied to the control atom; this pulse drives the excited state to the Rydberg state while it leaves the ground state unaltered. Then the 2$\pi$-pulse is applied to the target state as described above. This will lead to a change of phase of the qubit if the control atom is in the ground state while nothing occurs when it is in the Rydberg state. Finally, a $\pi$-pulse is applied again to the control atom to drive the Rydberg state to the excited state and thus to repopulate the computational levels. The Rydberg level is indeed only an ancillary level as Rydberg states would lie outside the Hilbert space spanned by the states of the computational basis. In such a way, the operator implemented is a controlled-$Z$ gate instead of the $ZZ$ gate presented in \eqref{eq:evolutionQAOA2}. $ZZ$, however, differs from the controlled-$Z$ gate just by single-qubit operators and a global phase. The mapping of all the parameters used in QAOA with Rydberg atoms is thus complete and can be used to create the desired state with Rydberg atoms.

\section{Cluster State and Error-Correction Encoding Circuit}
\label{sec: state and circuit}

The algorithm described in this work is general, and can even applied to any physical architecture for which the control Hamiltonians $H_X$ and $H_Z$ are applicable, but for an analysis of the results we focus on the success of the scheme for generating a 5-qubit full cluster state and for realizing the perfect error-correction encoding circuit using the Rydberg atom platform.
\subsection{Cluster State}
Cluster states were first introduced by Briegel and Raussendorf in 2001 \cite{PhysRevLett.86.910}. A general cluster state $\ket{C}$ is defined through the following eigenvalue equation: 
\begin{equation} 
K_a\ket{C}=(-1)^{k_a} \ket{C} ,
\end{equation} 
where the correlation operators are $K_a=X_a\otimes_{b\in N(a)}Z_b$, $N(a)$ identifies the neighbourhood of the qubit $a\in C$, and $k_a\in\{0,1\}$. 

A cluster state is a highly entangled state in terms of the Schmidt measure and, compared to other notorious quantum states such as GHZ and W states of the same size, it requires more projective measurements to be disentangled. This property, the so-called persistency of entanglement, is at the basis of one-way quantum computation with cluster states. In one-way quantum computation, a sequence of $m$ projective measurements is realized on a given subset of $m$ qubits of a $(n+m)$-qubit cluster state. The complementary subset of $n$ qubits that have not been measured will then define the $n$-qubit target state. This scheme, also known as measurement-based quantum computing, is realizable when initial states have an unbound entanglement width. In fact, this measure of bipartite entanglement is bound for most entangled states, including GHZ states, but it is unbound for cluster states. As such, the cluster state represents a precious resource for universal quantum computation so as to deserve the great attention it has attracted since being introduced. As an example, the implementation of this scheme for universal quantum computation has been proposed with Rydberg atoms. 

\begin{figure}[ht]
         \includegraphics[trim=160 500 20 50 ,clip,width=0.4\textwidth]{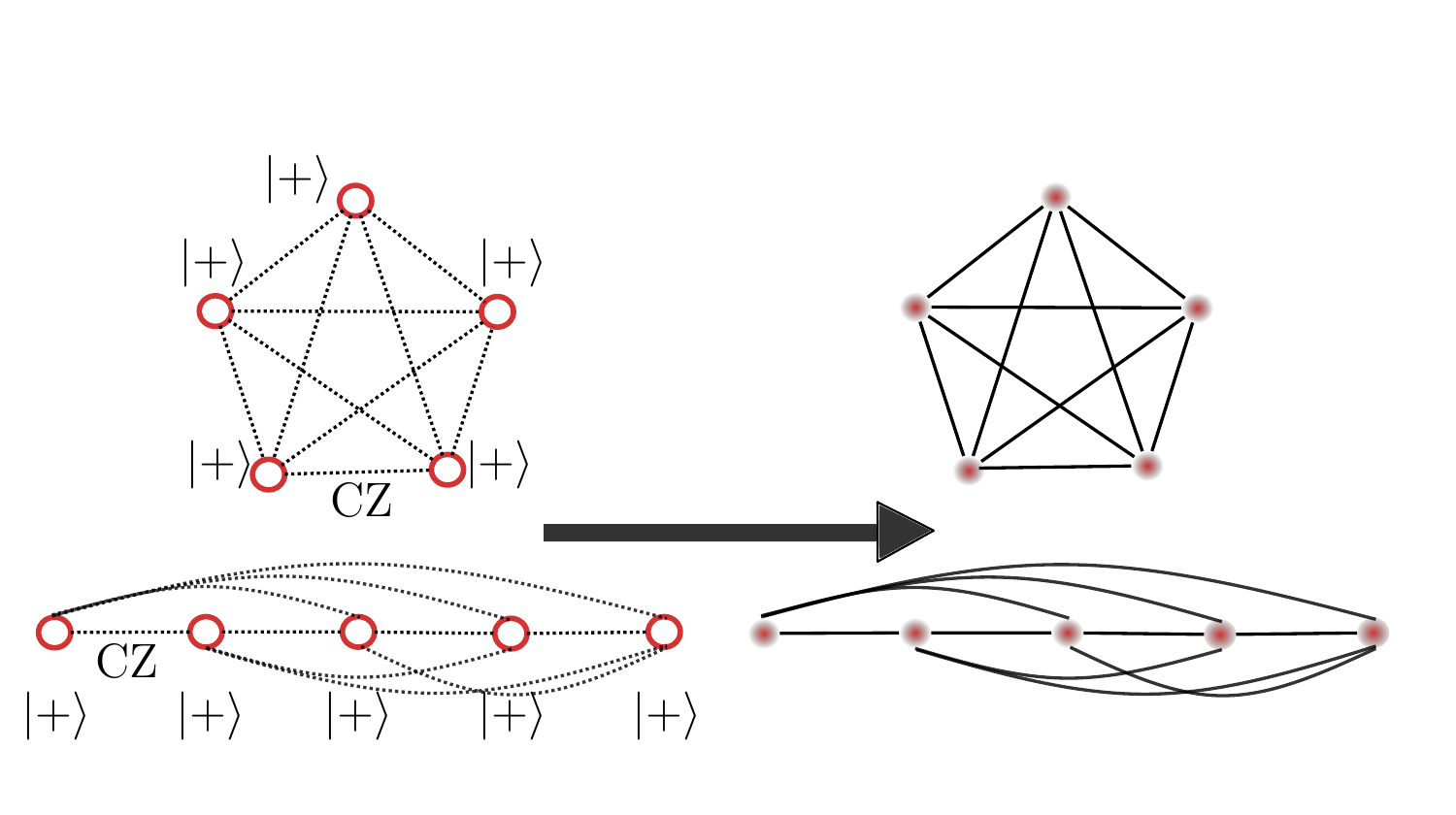}
         \caption{Diagram of ideal five-atom cluster state. Starting with all qubits (red circles) in the $\ket{+}$ state, the qubits are then entangled using a cZ gate to each adjacent (i.e. connected with a line) qubit within the diagram. Two separate geometries are depicted that realize the same cluster state.}
        \label{fig:ClusterStateDiagram}
\end{figure}

In order to test our scheme, we consider generating a fully entangled cluster state with $k_a=0$ and in which all the qubits are nearest neighbors of each other and as depicted in Fig.~\ref{fig:ClusterStateDiagram}.

\subsection{Error-Correction Encoding Circuit}

\begin{figure}[ht]
         \includegraphics[trim=60 370 380 387 ,clip,width=0.5\textwidth]{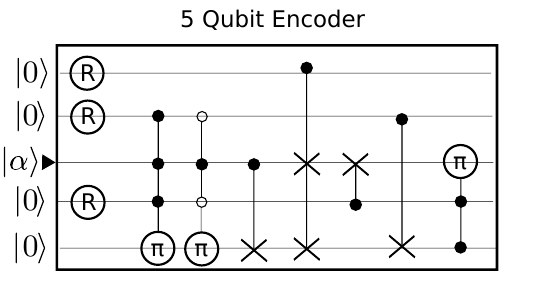}
         \caption{Perfect Quantum Error Correction gate \cite{PhysRevLett.77.198}, where the $i$th rows refers to the $i$th qubit. The circled R represents the application of a Hadamard gate, the circled $\pi$ represents a $\pi$ phase change, and the $\times$ represents a not gate. These gates are conditional upon all of the circles connected to them being satisfied, where a hollow circle implies the control state needing to be $\ket{0}$ while a filled circle is conditional on $\ket{1}$. The state being encoded is $\ket{\alpha}$ while the rest are auxillary qubits initialized in state $\ket{0}$.}
         \label{fig:ErrorGateDiagram}
\end{figure}

The efficacy of our algorithm is also evaluated by using it to directly implement the five-qubit circuit used to robustly encode quantum information in the Perfect Quantum Error Correcting Code \cite{PhysRevLett.77.198}. In this scheme, the quantum circuit reduces the errors on a logical qubit by distributing the information over five physical qubits: the qubit in the state one wants to preserve and four auxiliary qubits. Once the information of the qubit has been encoded in the five-qubit state, it is possible to retrieve it through a decoder circuit symmetric to the encoder. In fact, the states of the four auxiliary qubits will provide a diagnosis of the error witnessed by the system. Specifically, the code can detect both a sign and a spin flip of any of the qubits. The initial state can finally be restored with a simple unitary transformation of the main qubit.  A scheme of the error-correction encoding circuit is shown in Fig.~\ref{fig:ErrorGateDiagram}.

\section{Results}
\label{sec:results}

\subsection{Circuit depth}

\begin{figure}
\includegraphics[trim=35 355 30 180 ,clip,width=0.5\textwidth]{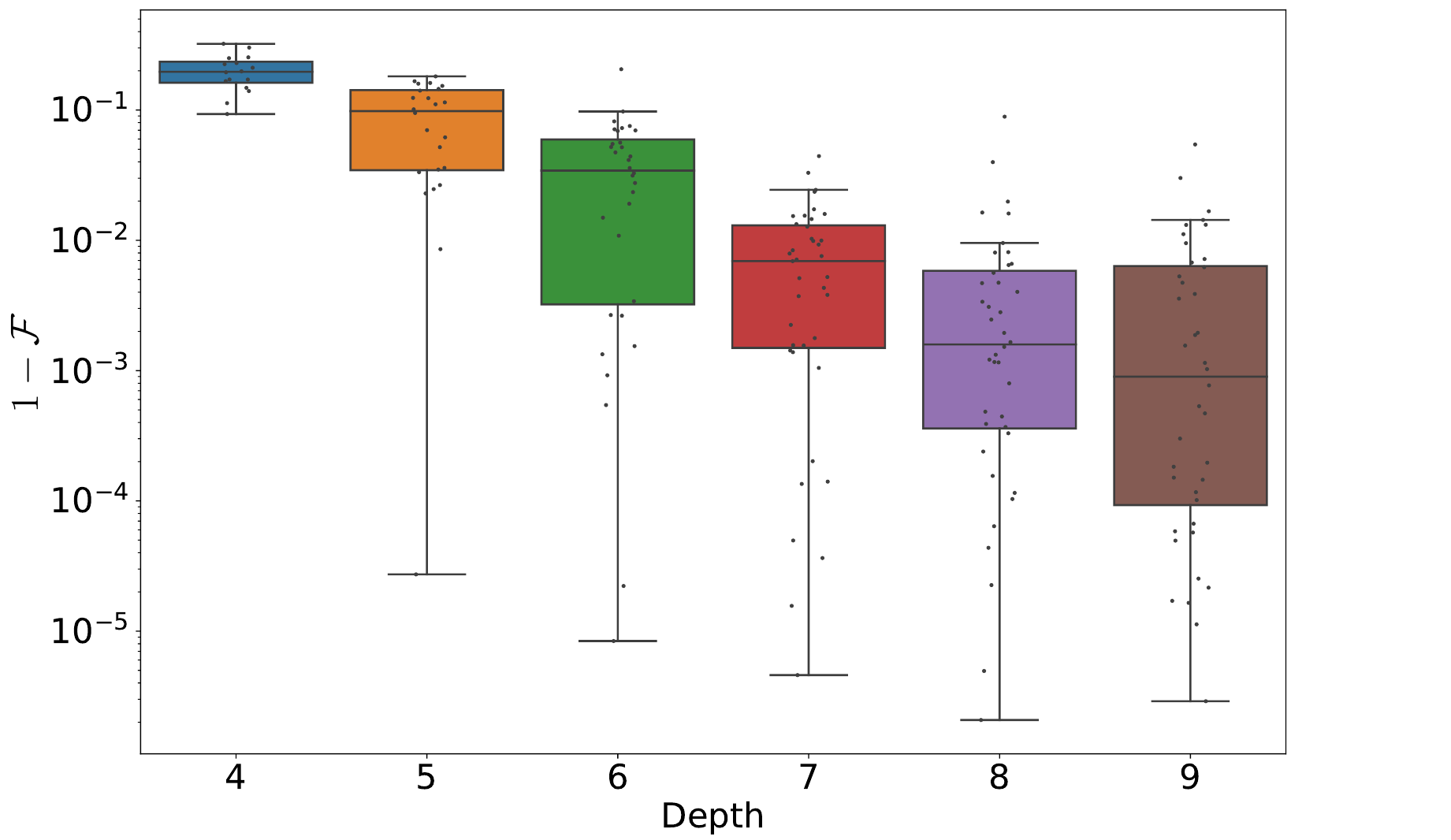}
\caption{One minus fidelities of the cluster states generated with QAOA after a dual annealing optimization of the parameters, plotted against the depth of the quantum algorithm. The semilogarithmic scale emphasizes the success of this method and how it grows with circuit depth.}
\label{fig:InfidelityCluster}
\end{figure}

\begin{figure}
\includegraphics[trim=35 355 30 180 ,clip,width=0.5\textwidth]{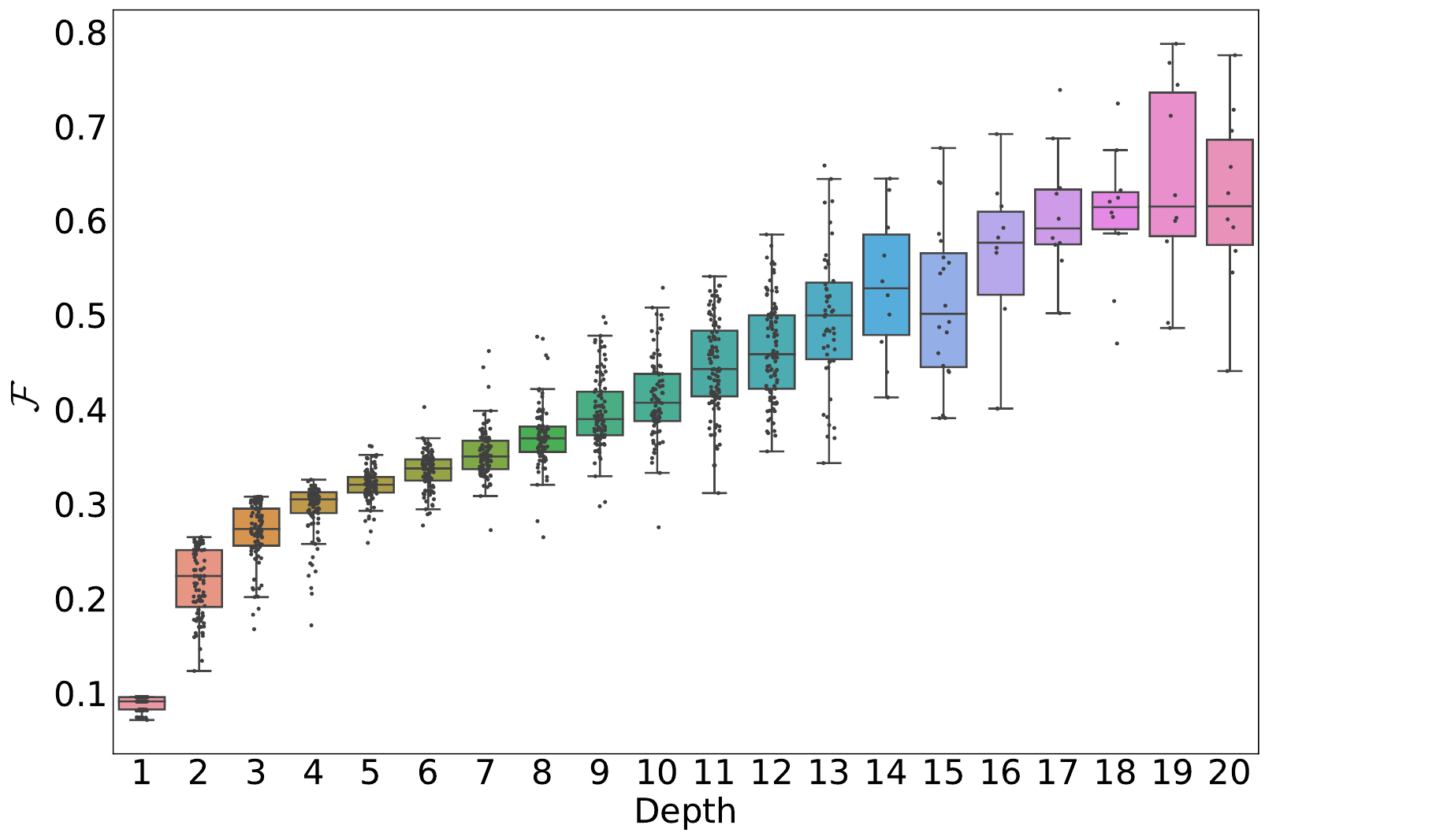}
\caption{Fidelities of the error-correction encoding circuit generated with QAOA after a dual annealing optimization of the parameters versus the depth of the quantum algorithm.}
\label{fig:FidelityErrorEncoding}
\end{figure}
The number of parameters used to prepare the target state or circuit is not fixed \textit{a priori} and the experimentalist chooses the one that provides satisfactory results on each occasion. We investigated how the quality of the algorithm depends on the depth of the circuit. 

First, we examine the abstract formulation shown in Eq.~\eqref{eq:evolutionQAOA2}. Optimizing the time parameters for this general formulation is in fact less time consuming than taking into account the Rydberg equivalences described in the previous section. Moreover, we observe that the Rabi angles found for the general case are good starting values for a subsequent optimization that incorporates details of the physical implementation.

The algorithm is tested by optimizing over different depths $p$. Once we have fixed the number of parameters $\mathcal{N}=5p$, we optimized them to find the state $|\psi_f\rangle$ generated by the algorithm that maximizes the fidelity  $\mathcal{F}=\left|\langle|\psi_f|C\rangle\right|^2$ with the target state $|C\rangle$. We sampled in such a way 40 states. We proceeded in an analogous way for different values of $\mathcal{N}$, by sampling 40 states for each $\mathcal{N}$. In particular, we consider a number of parameters ranging from 20 to 45 for the cluster state.
The optimization algorithm used in this analysis is the annealing optimization algorithm. It does not guarantee to provide the optimal set of parameters, but the fidelity of the state generated with the parameters provided by this optimization approximates the fidelity of the optimal state. In general, the optimization algorithm provides different results at each run, especially when the space of parameters has many dimensions. Therefore, averaging over 40 sampled states for each number of parameters improves the reliability of the results. A monotonic increase for the mean of the fidelities is observed by increasing the depth of the circuit.  In fact, the absolute value of the correlation coefficient between the the logarithm of the deviation from fidelity and the number of sequences is larger than 97\%. A plot of the results for the fidelities obtained for the cluster state and their dependence on the depth of the circuit is shown in Fig.~\ref{fig:InfidelityCluster}. Similarly, in Fig.~\ref{fig:FidelityErrorEncoding} we report the fidelities obtained for the error-correction encoding circuits generated by QAOA with different depths. In the case of the error-correction encoding circuit, the fidelity between unitary operators $V$ and $U$ has been defined as $\mathcal{F}=\abs{\Tr\left(V^\dagger U\right)}/\Tr\left(U^\dagger U\right)$.

Many time-consuming runs of the classical optimization algorithm are required to produce the plots reported in this work, with a significant number of parameter-dependent fidelities that are calculated at each depth. As a consequence, the plots show the results up to a given depth and do not necessarily show the best ones for the target circuit. However, higher depths are investigated with the even more time-consuming method that considers the Rydberg atoms implementation as described in the previous section. The quality of the prepared state witnesses an improvement with the depth of the algorithm in this case as well. The best results of the fidelities obtained for the error-correction encoding circuits increases linearly with a Pearson coefficient larger than 99\%, from a depth $p=18$, when $\mathcal{F}\simeq90.0\%$, until a depth $p=25$, when $\mathcal{F}=1$ is reached up to machine precision. 

Each point in the scatter plots of Figs. \ref{fig:InfidelityCluster} and  \ref{fig:FidelityErrorEncoding}  refers to a value of fidelity that has been evaluated as a result of an optimization of the Rabi angles for the cluster state and error-correction encoding circuit, respectively. Each optimization provides a different result because of the intrinsic randomness of the optimization algorithm that has been used. Therefore, the optimization result is not necessarily the best result possible at the given depth. 
In order to provide the reader with a measure of the behaviour of the optimization algorithm and of the probability of getting a given result with it, we have chosen to report all of the evaluated fidelities with overlain scatter and box plots.  
As an alternative to this representation, we pondered realizing the extreme value analysis of the collected results. The extreme value statistics would have provided the probability density of obtaining a given fidelity value after a fixed number of optimizations. In order to do so, we need to identify the probability density that fits the observed distribution of data around the expected maximum.  
However, this identification proves to be prohibitive in our case. The maximum reachable fidelity at each depth has not been identified here, with the results providing just a minimum for it. Moreover, even fixing a value for this maximum, the number of recorded fidelities is not large enough to determine with satisfactory accuracy the fitting parameters. 

\begin{figure}
\includegraphics[trim=30 0 0 0 ,clip,width=0.5\textwidth]{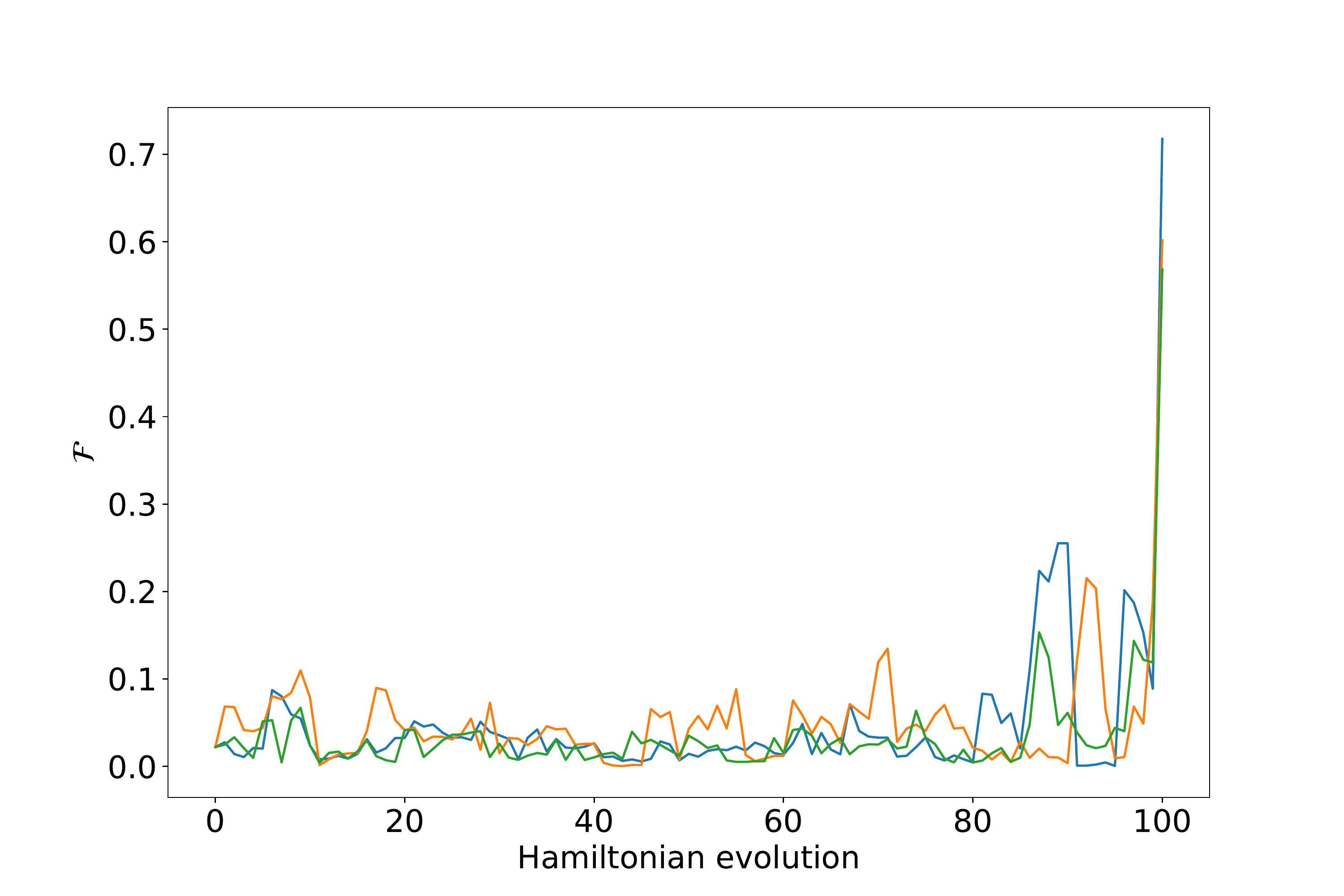}
\caption{Plot of the fidelities with the target error-correction encoding circuit and the operation realized after a given Hamiltonian evolution for three instances of QAOA of depth $p=20$. In general, the fidelity is far from its final value until the last step.}
\label{fig:FidelitiesEvolution}
\end{figure}

\begin{figure}
\includegraphics[trim=30 0 0 0 ,clip,width=0.5\textwidth]{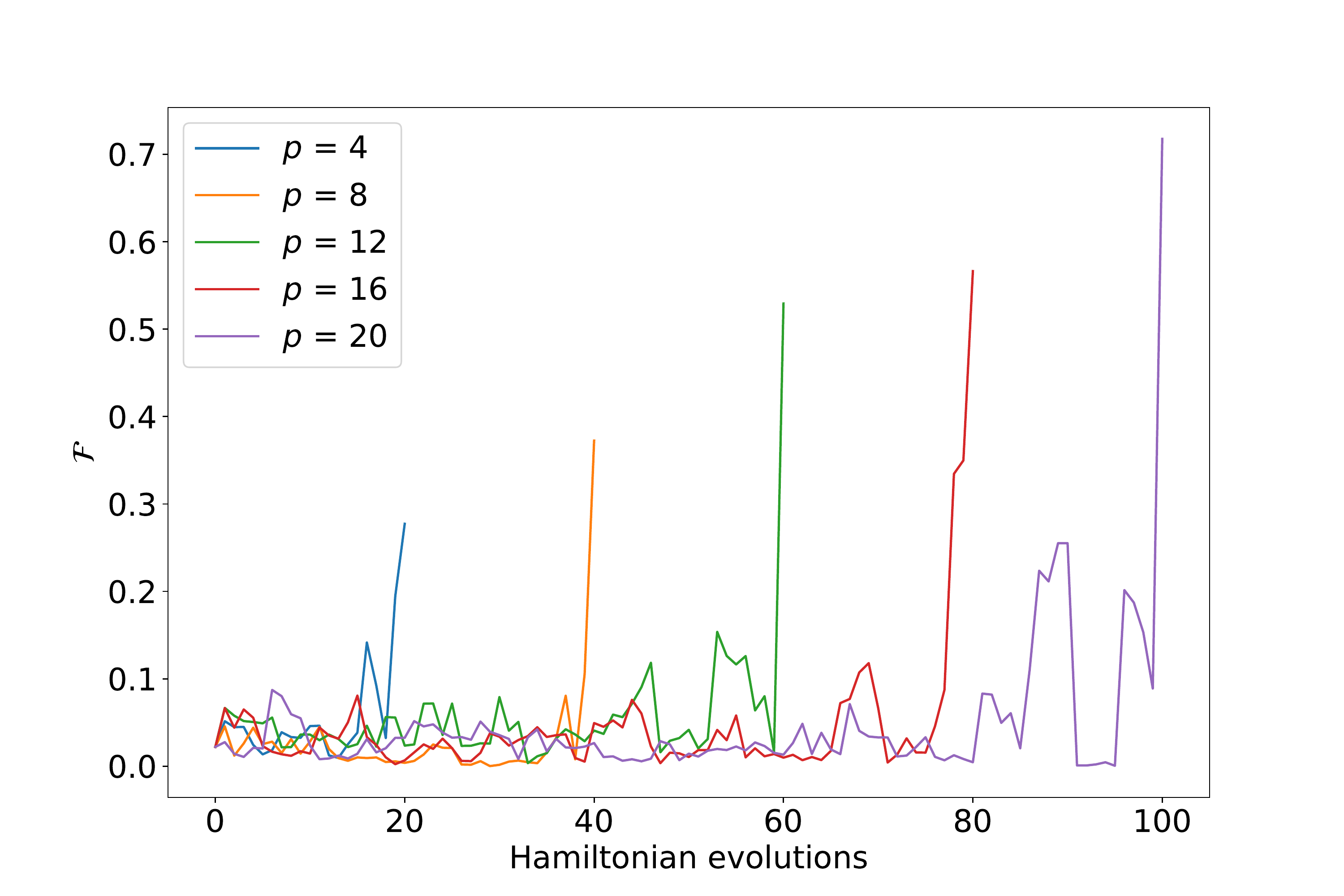}
\caption{Plot of the fidelities with the target error-correction encoding circuit and the operation realized after a given Hamiltonian evolution for 5 instances of QAOA of depth $p=4, 8, 12, 16$ and $20$. In general, the fidelity is far from its final value until the last step. }
\label{fig:FidelitiesEvolutionDepths}
\end{figure}

To gain more insight into the optimization procedure, the plot in Fig.~\ref{fig:FidelitiesEvolution} displays three examples of the evolution of the fidelity with the perfect Quantum Error Correcting Code after each Hamiltonian evolution of the algorithm with a depth of 20. With the expression "Hamiltonian evolution" we refer to each of the terms introduced in Eq.~\eqref{eq:evolutionQAOA2}. We observe that the fidelity does not increase steadily towards the final value but instead experiences an abrupt jump at the last step for each of the three examples. This suggests that fidelity might not be the best measure to capture the path of the state towards the target. Analogously, the plots for instances with depth equal to 4, 8, 12, 16, and 20 are compared in Fig.~\ref{fig:FidelitiesEvolutionDepths}. We observe that the states have unique fidelity trajectories for different depths, evolving along paths through Hilbert space that are nonmonotonic in any distance measure induced by the fidelity. Importantly, the optimal results for longer-depth circuits are not obtained by adding fine-tuning steps after the optimal results of the shorter-depth circuits; each trajectory is unique. These counter-intuitive paths demonstrate the usefulness of relying on our algorithm to find optimal control operations.

\begin{figure}
\includegraphics[trim=0 0 0 0 ,clip,width=0.45\textwidth]{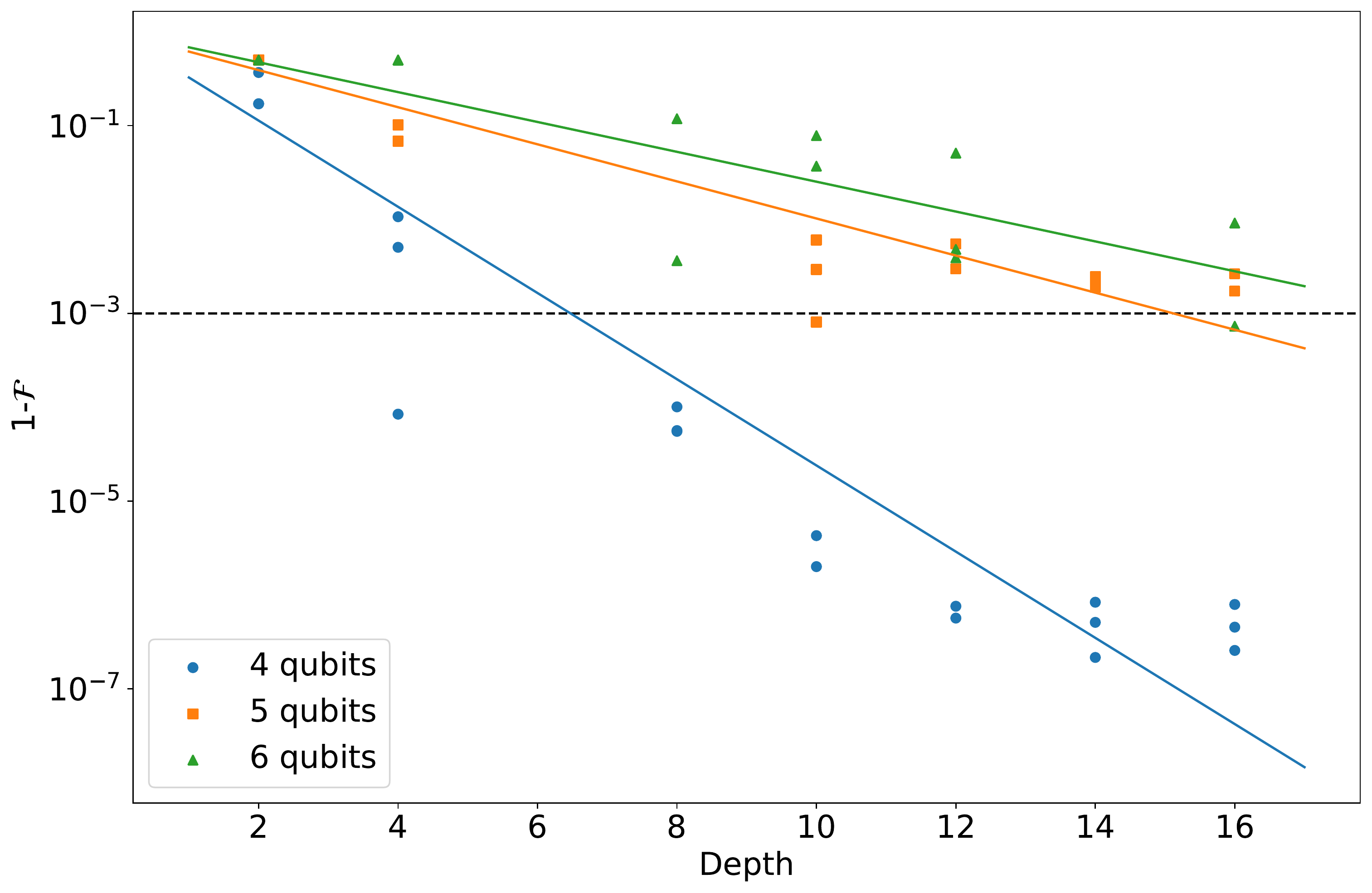}
\caption{Fidelities obtained by simulating control on 4, 5, and 6 Rydberg atoms to create 4, 5, and 6-qubit fully connected cluster states using QAOA algorithms of different depths. The dots represent the classical optimizations run at each depth for 4, 5, and 6-qubit cluster state; the straight lines are the exponential fit of the observed data. Fixing one minus the fidelity of the initial $n$-qubit state to $a_n$, the parameters $\lambda_n$ of the fitting function $a_ne^{-\lambda_np}$ take the following values, $\lambda_4\simeq1.06, \lambda_5\simeq0.46,$ and $\lambda_6\simeq0.37$. The line at $\mathcal{F}=99.9\%$ highlights how systems with a smaller number of qubits reach this fidelity value at a lower depth. }
\label{fig:FidelitiesCluster456}
\end{figure}

In order to evaluate how the fidelities at given depths scale with the number of qubits of the target state, we analyze the trend of fidelities obtained by simulating control on 4, 5, and 6 Rydberg atoms as described in section~\ref{sec:equivalence} to realize 4, 5, and 6-qubit cluster states, respectively. The results are plotted in Fig.~\ref{fig:FidelitiesCluster456}.
We observe that at a given depth, target states with less qubits can be realized with higher fidelity. We can conclude that for a given threshold fidelity, larger depths are required to realize larger qubit states. As an example, a fidelity of $99.9\%$ is reached at depths of $p=4, 10$, and 16 for the 4, 5, and 6-qubit cluster states, respectively.

\subsection{Robustness}
In the analysis made so far, no error has been introduced in the calculations, but errors cannot be neglected in real-life experiments with Rydberg atoms. The lifetime of the Rydberg atoms and the Rydberg shift in the blockade are indeed the main features affecting the fidelity of the results. Therefore, a careful choice of the intensities of the pulses is crucial to increase the fidelity of the results \cite{PhysRevA.85.042310}. On one hand, each intensity needs to be large enough to allow short time pulses while, on the other, we want to limit the off-resonant transitions as best as possible that arise from too-strong pulses. As such, we investigate how the fidelity of the generated state and, therefore, the optimal depth of the circuit are affected by the presence of some noise in the system. 

In our theoretical model, we incorporate all of the errors from different sources into deviations from the optimal parameters of QAOA found for the noiseless system. To this end, we exploited the periodicity over $2\pi$ of the arguments of the unitary operators in Eq. \eqref{eq:evolutionQAOA2}. Indeed, a different value randomly picked from a uniform distribution in the interval $[-\pi R,\pi R]$ is added to each parameter of every optimal sequence. The coefficient $R$ is referred to as the magnitude of the error. When $R$ is equal to zero, no change is made to the sequences of parameters and when $R$ is equal to 1 the parameters turn out to be uniformly distributed in an interval of length $2\pi$, thus losing any information about the target state. Finally, for each sequence we calculate the fidelity of the state produced by the new set of parameters. This procedure is repeated many times for each state in order to average over the statistical fluctuations of the simulated noise.

\begin{figure}[H]
\includegraphics[trim=35 355 30 180 ,clip,width=0.5\textwidth]{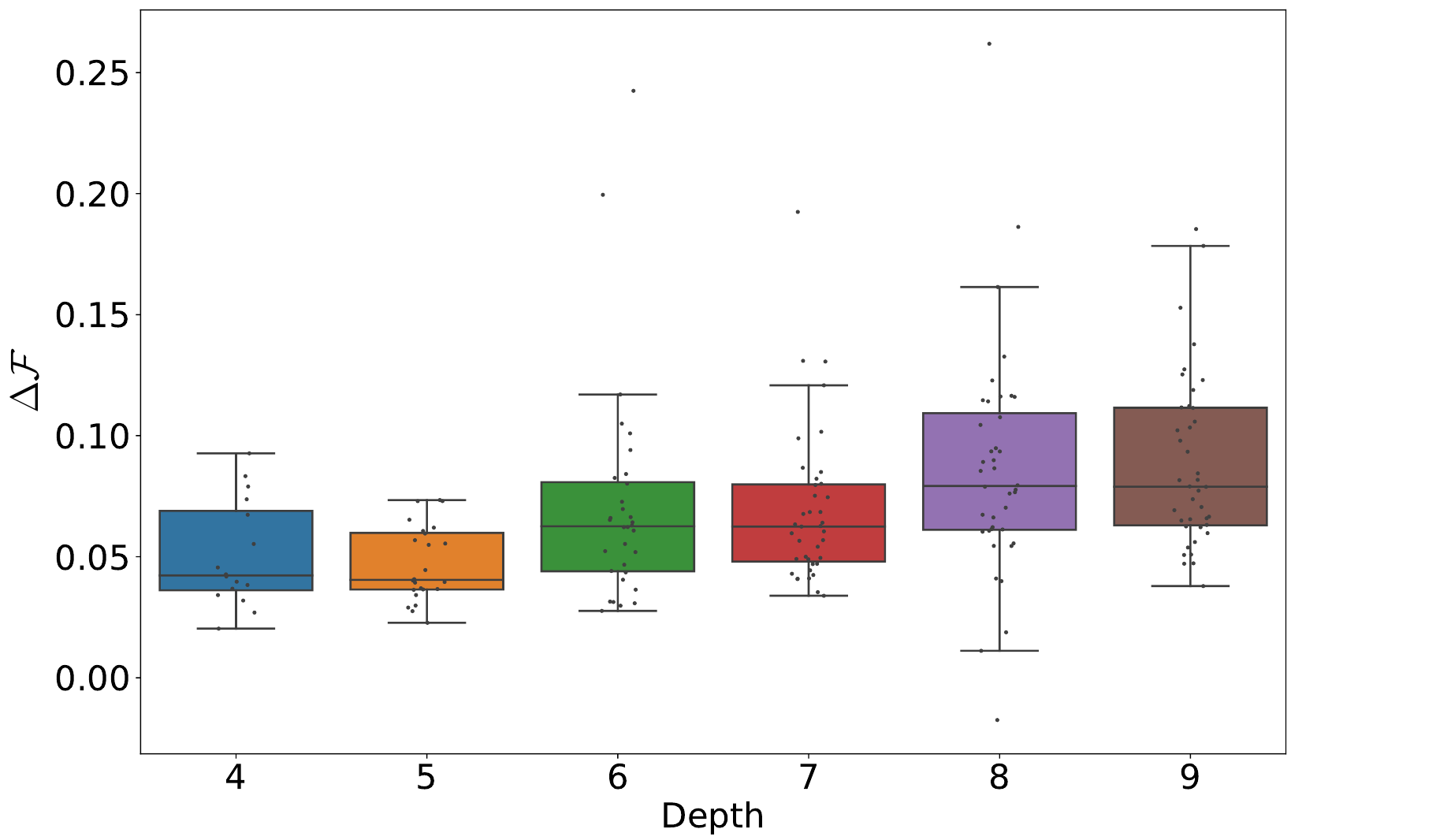}
\caption{Decreases in the fidelities of the cluster states generated with QAOA after a dual annealing optimization of the parameter with noise level $R=1\%$ relative to the same optimization without noise, plotted against the depth of the quantum algorithm.}
\label{fig:DeltaClusterState}
\end{figure}
\begin{figure}[H]
\includegraphics[trim=35 355 30 180 ,clip,width=0.5\textwidth]{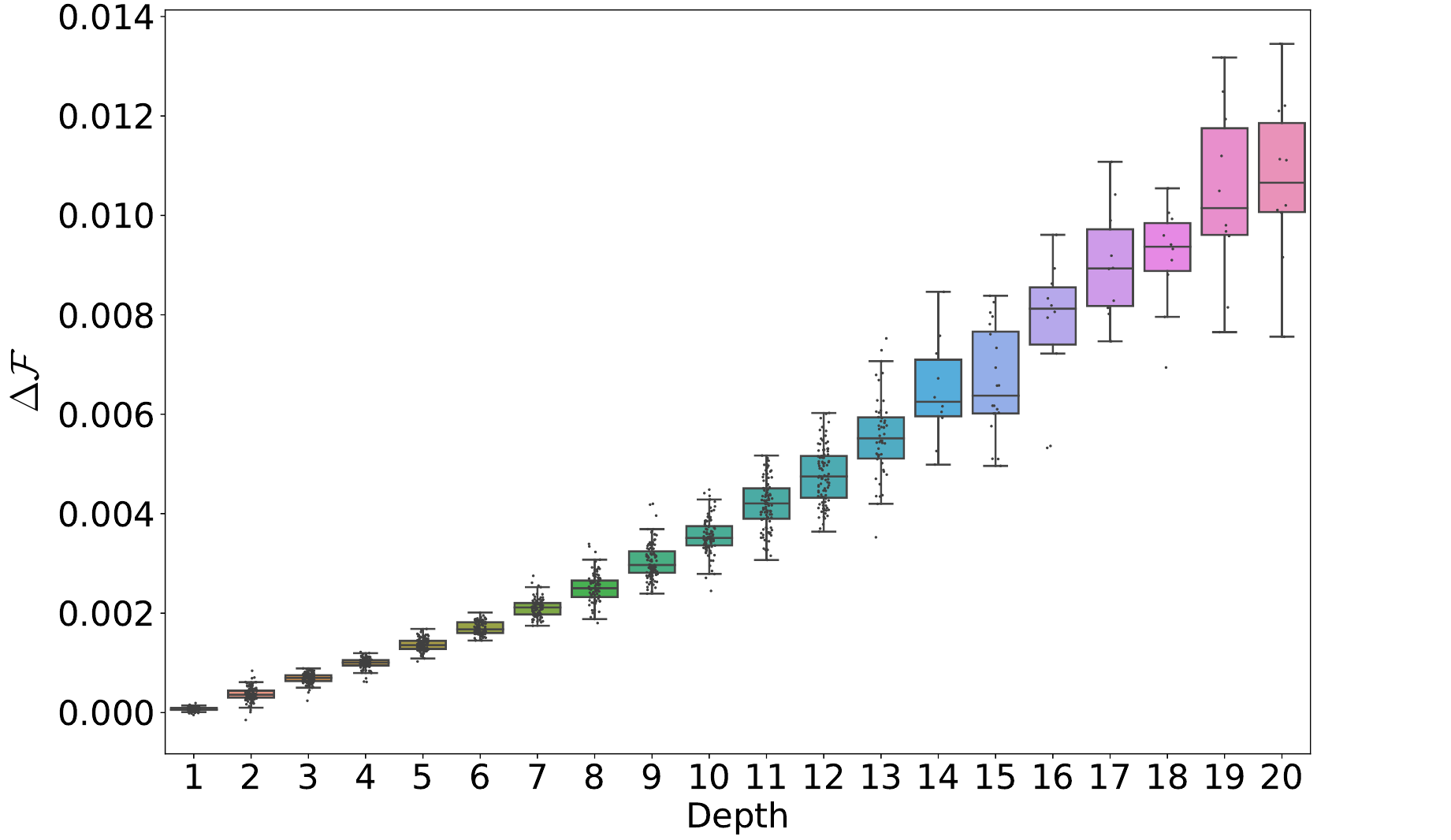}
\caption{Difference between the fidelities of the error-correction encoding circuit generated with QAOA after a dual annealing optimization of the parameter without noise and with noise level $R=1\%$ against the depth of the quantum algorithm.}
\label{fig:DeltaErrorEncoding}
\end{figure} 

The differences between the results obtained with no noise and after introducing a noise level $R=1\%$ are reported for the cluster state and the error-correction encoding circuit in Figs. \ref{fig:DeltaClusterState} and \ref{fig:DeltaErrorEncoding}, respectively as a function of the depth of the circuit.
For both applications, we observe a loss of fidelity that increases with the depth of the circuit. This behaviour suggests the existence of a threshold depth for a given noise level that maximizes the quality of the generated state. The depths considered in this work, however, do not allow us to identify such a threshold value; one might be found by investigating the results obtained with deeper circuits. 
The values of fidelities with $R=0,\, 10^{-3},$ and $10^{-2}$ reachable by the QAOA protocol for Absolutely Maximally entangled states with 5 (AME5) and 6 (AME6) qubits, and for GHZ state are shown in table~\ref{tab:FidelityOhterStates}.

\begin{table}[h]
\begin{tabular}{| c | c | c | c |}
\hline
State & Fidelity & R = 1\%  & R = 0.1\% \\ 
\hline
 GHZ & 99.93\% & 97.7\% & 99.91\% \\ 
 \hline
 AME5 & 99.9\% & 98.5\% & 99.86\% \\ 
\hline
 AME6 & 99.83\% & 98.3\% & 99.82\% \\
 \hline
\end{tabular}
\caption{Fidelity of the generated states with and without noise.}
\label{tab:FidelityOhterStates}
\end{table}

\subsection{Physical implementation}

The physical implementation considered in the simulation is mostly based upon the experimental setup proposed in \cite{Bernien2017}. The isotopes of $^{87}$Rb that are used as Rydberg atoms, are arranged in a linear array of trough optical tweezers. The spacing is set to $5.24\,\mu$m, which results in a nearest-neighbour interaction strength of $V_{i,i+i}=2\pi\times24\,$MHz. The equivalence between the logical and the Rydberg states and the energetic levels of the atoms are $\ket{0}=\ket{5S_{1/2},F=2,m_F=-2}$, $\ket{1}=\ket{6SP_{13/2},F=3,m_F=-3}$, and $\ket{r}=\ket{70S_{1/2},J=1/2,m_J=-1/2}$.
The frequencies of the driving pulses from the ground to the excited state $\Omega_b$, and from the excited to the Rydberg state $\Omega_r$ are $\Omega_b=2\pi\times60\,$MHz, and $\Omega_r=2\pi\times36\,$MHz.

\begin{figure}
\includegraphics[width=0.49\textwidth]{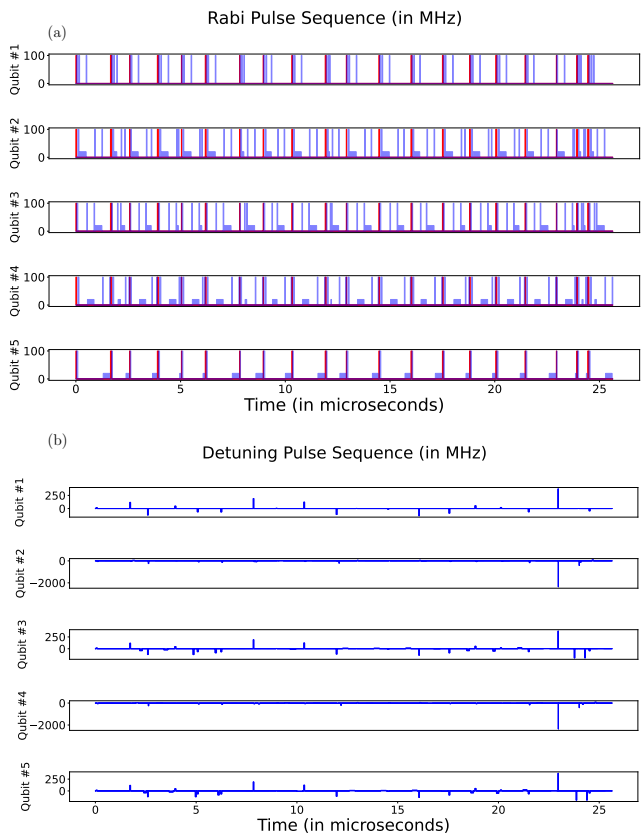}
     
\caption{(a) Rabi frequency pulse sequence for the generation of the error-correction encoding circuit. The pulse sequence consists of a series of square shaped pulses acting upon the different qubits 1-5, where the red pulses refer to driving fields from $\ket{0} \mapsto \ket{1}$ and the blue pulses refer to driving fields from $\ket{1} \mapsto \ket{r}$. (b) Frequency detuning $\Delta$ of the blue pulses. }
        \label{fig:PulseSequences}
\end{figure}

With these numbers, one can use the algorithm to construct a set of pulse sequences required to create any state or enact any quantum circuit. As an example, the pulses required to prepare a five-qubit error-correction encoding circuit are reported in Fig. \ref{fig:PulseSequences}.
The intensity of the laser illuminating the target qubit, in the sequences of pulses that realize the controlled-$Z$ operators, has been set to be smaller than the intensity used for all the other pulses so as to minimize the possibility of off-resonant transitions when the Rydberg blockade is induced by the control qubit.

The state $\ket{e}$ is not long-lived compared to the duration of the sequence or the ground ($\ket{g}$) and the Rydberg ($\ket{r}$) states. This necessitates using an ancillary pulse to coherently shuffle population between the intermediate state ($\ket{e}$) and an additional long-lived ground spin state of the Rb atom. This is routinely done in quantum memory experiments with atomic ensembles.

\section{Discussion and Conclusions}

The quantum circuit model is the most common model of quantum computation. It makes use of multi-qubit states on which the information is encoded, processed, and extracted, and quantum circuits that realize these operations. Therefore, quantum state and circuit generation are crucial prerequisites for an efficient quantum computation. 
As an example, entangled states are an especially precious resource for quantum computation without any classical counterpart. But their preparation turns out to be, in general, a complex task that might require significant resources to be realized \cite{Li2021}.
Our state-generation scheme with QAOA aims to overcome these challenges by employing a classical optimization of the parameters and a widely used physical implementation.

The universality of QAOA is the main cue for the realization of universal quantum control with this algorithm. The protocol proposed in this work guarantees that any $n$-qubit state can be achieved with the desired approximation just by using local and two-local interactions among qubits. The depth of the circuit and the interaction times are determined according to the target state and the fidelity that the experimenter wants to reach. The theoretical description of the scheme and the efficiency in its abstract formulation are independent of the possible implementation. Only once a physical setup is chosen for the implementation of the scheme, then a proper, eventually non-trivial, mapping of the parameters in terms of the system Hamiltonian must be taken into account. An array of Rydberg atoms, where the two-atom interaction is realized through the Rydberg blockade phenomenon is chosen for its popularity.

The results achieved for the quantum states taken into consideration show that quantum control with QAOA can be a valid scheme for many applications. The efficacy of the scheme appears more clearly when one wants to initialize more complex states, for example with a higher entanglement, and where other protocols require a significant amount of resources. Indeed, it is less affected by the state that must be initialize; it operates on a widely employed implementation that requires just nearest neighbor interactions and a limited number of pulses to produce highly entangled quantum states and even a five-qubit quantum circuit with a significant number of gates. 
The fidelity of the states and circuits might be further increased through a more reliable optimization of the parameters realizable with more computation power and advanced optimization protocols. 
The simulation that supports the findings of this study is openly available on GitHub~\cite{Github}.


\begin{acknowledgments}
    The authors acknowledge that the NRC headquarters is located on the traditional unceded territory of the Algonquin Anishinaabe and Mohawk people. This work was supported by the Natural Sciences and Engineering Research Council of Canada (NSERC) through its Discovery Grant (DG) program and Postdoctoral Fellowships (PDF) program.
\end{acknowledgments}

\end{document}